# Cyberattack Detection using Deep Generative Models with Variational Inference

Sarin E. Chandy; Amin Rasekh; Zachary A. Barker; M. Ehsan Shafiee

Xylem Inc.

**Abstract.** Recent years have witnessed a rise in the frequency and intensity of cyberattacks targeted at critical infrastructure systems. This study designs a versatile, data-driven cyberattack detection platform for infrastructure systems cybersecurity, with a special demonstration in water sector. A deep generative model with variational inference autonomously learns normal system behavior and detects attacks as they occur. The model can process the natural data in its raw form and automatically discover and learn its representations, hence augmenting system knowledge discovery and reducing the need for laborious human engineering and domain expertise. The proposed model is applied to a simulated cyberattack detection problem involving a drinking water distribution system subject to programmable logic controller hacks, malicious actuator activation, and deception attacks. The model is only provided with observations of the system, such as pump pressure and tank water level reads, and is blind to the internal structures and workings of the water distribution system. The simulated attacks are manifested in the model's generated reproduction probability plot, indicating its ability to discern the attacks. There is, however, need for improvements in reducing false alarms, especially by optimizing detection thresholds. Altogether, the results indicate ability of the model in distinguishing attacks and their repercussions from normal system operation in water distribution systems, and the promise it holds for cyberattack detection in other domains.

## 1. Introduction

Industrial control systems (ICS) are computers that control operation of industrial processes, from centralized water treatment plants and oil refineries to distributed water distribution systems and power grids. ICSs are an integral part of modern critical infrastructure systems and may include supervisory control and data acquisition (SCADA), distributed control systems, and other smaller types of control configurations such as programmable logic controllers (Stouffer et al. 2015).

Water and Wastewater Systems Sector is one the sixteen sectors officially designated as Critical Infrastructure Sectors in the United States (White House 2013). Water infrastructures have been traditionally standalone systems with little or no cyber connections. The burgeoning wet infrastructure modernization, however, has lead into merging of information technology and operation technology in infrastructure. This integration accountable for such modernization and its benefits, has also exported the malignancies of cyberspace to water infrastructure and their ICSs, introducing new risks. This is evidenced by multiple real incidents such as the Tijuana River sewage spill in 2012 and recent hack of unnamed water utility, which lead to unauthorized manipulation of flow and disinfection controllers (Rasekh et al. 2016; Verizon Communications 2016). In 2013, hackers gained unauthorized remote access to the SCADA systems of a small dam in the State of New York, enabling them to acquire information about water levels and the status of the sluice gate (Thomson 2016). Water sector has been recognized as one of the highest-targeted sectors amongst the critical infrastructure sectors by the U.S. Department of Homeland Security (ICS-CERT 2016) and strengthening them against cyber threats is a national concern and priority (White House 2017).

While the conversation about the cybersecurity of water resources systems has been going on for long (Water Sector Coordinating Council 2008; Cardenas et al. 2009), its reception and recognition by the water research community is relatively new. Two pioneering studies by Amin et al. (2013a; 2013b) performed simulation-based assessment and detection of stealthy deception attacks launched against cascaded canal systems. Perelman et al.

(2014) formulated the network vulnerability as an attacker-defender problem to identify the most critical links in a network. Rao and Francis (2015) conducted a descriptive, critical review of cybersecurity practices and risk management in the drinking water distribution sector. Laszka et al. (2017) provided theoretical foundations for synergizing redundancy, diversity, and hardening for resilience against cyber-physical contamination attacks. Housh and Ohar (2017) and Ahmed et al. (2017) developed model-based cyberattack detection tools for water distribution systems. Taormina et al. (2017) developed a modeling framework, named epanetCPA, which quantitatively characterizes the effect of attacks on the hydraulic behavior of water distribution systems. This article, along with a recent editorial in this journal (Rasekh et al. 2016), provide a detailed, informative review of water infrastructure cybersecurity evolution, concepts, and literature.

Cyberattacks to ICSs often manifest themselves as anomalous behaviors that do not conform to the norm of the system. One exception might be eavesdropping attacks because the system operations would not be altered by such attacks. Manual detection of anomalies is extremely laborious and difficult, if not impossible, given the extreme intricacy of the modern distributed systems (Chandola et al. 2009). Simulation-based and conventional machine-learning approaches also have limited versatility due to their inherent dependency on substantial domain knowledge (Amin et al. 2013a; 2013b) and inability to process data in their raw form (LeCun et al. 2015). In this article, we design a variational autoencoder (VAE) (Kingma and Welling 2014), which is a deep generative model with variational inference, for detection of cyberattacks. This study pioneers the formulation and application of VAE for cyberattack detection.

In what follows, a description of the VAE-based detection methodology and its algorithmic theory is first provided. Next, the evaluation criteria used to assess the detection performance are presented. This is followed by a description of data requirements and experiment setup for detection of simulated cyberattacks to an illustrative water distribution network system – C-Town. Implementation results and discussions are provided after. This article concludes with an informative series of extension ideas for future studies.

## 2. Methodology

VAE is a deep generative model with variational inference. It mines the heterogeneous ICS data collected from a variety of sensors that are normally included in a modern SCADA. The VAE processes data in its raw form to learn and reproduce behavior of the underlying system. This section describes the fundamental concepts and elaborates on the main engine of our detection methodology.

### 2.1. Deep Learning

Deep learning combines several simple, nonlinear processing layers to discover and learn complex representations of raw system data. While the concept of deep learning has existed for decades, its rise to prominence goes back only a few years, sparked arguably by the astonishing performance of convolutional neural networks in the ImageNet contest in 2011 (Krizhevsky et al. 2012). Ever since, it has stimulated breakthroughs in many domains from computer vision to genomics (LeCun et al. 2015). Deep learning has been also utilized in the water resources domain recently for precipitation estimation (Tao et al. 2016), reservoir inflow forecasting (Bai et al 2016), and hydrological inference (Marcais and Dreuzy, 2017). Deep learning holds promise for detection of cyberattacks to ICSs as well given the heterogeneity and massiveness of their data and intricacy of the underlying systems.

### 2.2. Generative Models

Generative models are a branch of unsupervised learning techniques in deep learning that can learn a phenomenon by analyzing a large number of existing samples from it and then generate new samples like it (Hinton and Ghahramani, 1997). Since a generative model's number of parameters is substantially smaller than the amount of data it is trained on, it is forced to discover and embody the essence of the data to generate data like it. Generative Adversarial Networks (Goodfellow et al. 2014), Deep Autoregressive Networks (Gregor et al. 2014), and Variational Autoencoders (Kingma and Welling 2014) are amongst the most popular and powerful examples of generative model approaches. The application of a Variational Autoencoders (VAE)-based model is evaluated here to detect cyberattack using the ICS data that represents the performance of water distribution network.

### 2.3. Variational Autoencoder

A VAE is a generative model that works by optimizing a variational lower bound of the likelihood of the data. As a hybrid of deep learning and variational inference, VAE was introduced to perform efficient learning in the presence of intractable posterior distributions and large datasets. Variational inference is a method that approximates probability densities via optimization (Blei et al. 2016). A variational lower bound is a lower bound of the likelihood of the observed data. Thus, a higher value for the lower bound will imply a model with a better fit to the observed data. Using a lower bound approximation of the likelihood enables the VAE to perform efficient learning.

A VAE has two halves: an encoder and a decoder. The encoder learns a latent representation of the data, and the decoder learns to convert this representation back into the original data. The two recognition and generative halves are jointly trained by maximizing probability of the input data.

The data are often observations of the system over a long period of time. In the context of water distribution systems, this may include time series of tank water level, nodal pressure, pump status, and similar. The observational data are acquired by the sensors installed on the different components of a water distribution system. Nevertheless, the VAE strategy developed and deployed here is fully generic and functions independent of any internal working of the system it is applied to and the nature of the observations data fed into it. The model developed here can be taken as is and applied to any other system for which observations over a long period of time are available.

We provide a quick derivation of the variational lower bound in the context of our application here. A given data (e.g., time series of pressure at the outlet of a pump station) is represented by the variable $\boldsymbol{x}$ while the encoding is represented by the variable $\boldsymbol{z}$. The encoder network with parameters $\phi$ encodes the given input data $\boldsymbol{x}$ as $\boldsymbol{z}$ with the distribution given by $\boldsymbol{q_{\phi}(z|x)}$ while the decoder network with parameters $\theta$ decodes $\boldsymbol{z}$ into $\boldsymbol{x}$ with the probability $\mathbf{p_{\theta}(x|z)}$. The encoder tries to approximate the true but intractable posterior produced by the decoder represented as $\boldsymbol{p_{\theta}(z|x)}$. By assuming a standard normal prior (i.e., $\boldsymbol{z} \sim N(0,1)$) for the decoder, we can optimize the network parameters by maximizing the probability of the data $\boldsymbol{p_{\theta}(x)}$.

Latent representation of the data, which is defined by latent variables $\boldsymbol{z}$, is a simpler and lower-dimensional representation of the high-dimensional input data (Bengio et al. 2013). The latent space internalizes the essence of the data, which is discovered through training, to generate data like the input high-dimensional data it was trained on. This process is schematically illustrated in Figure 1. In the context of water networks, a high-dimensional raw data point can be a 24-hour window of a set of time series generated by the installed pressure, flow, and other sensors.

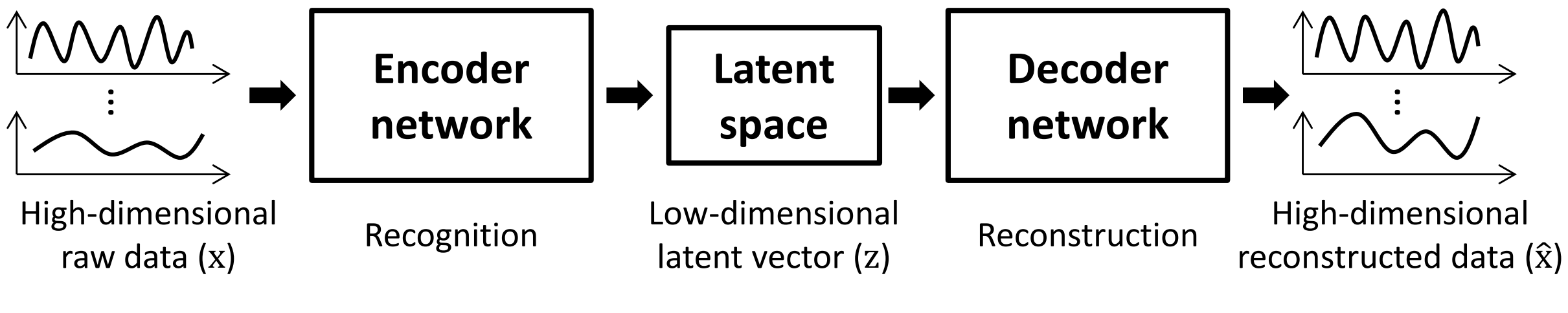

**Figure 1.** The encoder learns a low-dimensional, latent representation of the data, and the decoder learns to reconstruct this representation back into the original data. The recognition and reconstruction halves are jointly trained to discover and embody the essence of the data.

Let the log-likelihood of the data be

$$log\ p(\boldsymbol{x}) = log\ p\left(\boldsymbol{x}^{(\mathbf{1})}, \dots, \boldsymbol{x}^{(\boldsymbol{n})}\right) = \sum_{i=1}^{n} \log p(\boldsymbol{x}^{(\boldsymbol{i})}) \tag{1}$$

Therefore,

$$\log \boldsymbol{p_\theta}(\boldsymbol{x}) = D_{KL}(\boldsymbol{q_\phi}(\boldsymbol{z}|\boldsymbol{x}^{(\boldsymbol{i})})||\boldsymbol{p_\theta}(\boldsymbol{z}|\boldsymbol{x}^{(\boldsymbol{i})})) + \mathrm{E}_{\boldsymbol{q_\phi}(\boldsymbol{z}|\boldsymbol{x}^{(\boldsymbol{i})})}\left[-\log \boldsymbol{q_\phi}\left(\boldsymbol{z}|\boldsymbol{x}^{(\boldsymbol{i})}\right) + \log \boldsymbol{p_\theta}(\boldsymbol{x}, \boldsymbol{z})\right] \tag{2}$$

where $D_{KL}$ is the KL Divergence. Since it is always non-negative, we arrive at

$$\log \boldsymbol{p_\theta}(\boldsymbol{x}) \geq \mathrm{E}_{\boldsymbol{q_\phi}(\boldsymbol{z}|\boldsymbol{x}^{(\boldsymbol{i})})}\left[-\log \boldsymbol{q_\phi}\left(\boldsymbol{z}|\boldsymbol{x}^{(\boldsymbol{i})}\right) + \log \boldsymbol{p_\theta}(\boldsymbol{x}, \boldsymbol{z})\right] = \mathcal{L}(\theta, \phi; \boldsymbol{x}^{(\boldsymbol{i})}) \tag{3}$$

where $\mathcal{L}\left(\theta, \phi; \boldsymbol{x}^{(\boldsymbol{i})}\right)$ is the variational lower bound, and can be rewritten as

$$\mathcal{L}\left(\theta, \phi; \boldsymbol{x}^{(\boldsymbol{i})}\right) = -D_{KL}(\boldsymbol{q_\phi}(\boldsymbol{z}|\boldsymbol{x}^{(\boldsymbol{i})})||\boldsymbol{p_\theta}(\boldsymbol{z})) + \mathrm{E}_{\boldsymbol{q_\phi}(\boldsymbol{z}|\boldsymbol{x}^{(\boldsymbol{i})})}\left[-\log \boldsymbol{p_\theta}\left(\boldsymbol{x}^{(\boldsymbol{i})}|\boldsymbol{z}\right)\right] \tag{4}$$

In order to reduce variance of this estimate, a reparametrization trick is applied, as fully described in Kingma and Welling (2014). Given the intricacy of the underlying ICS, we use a deep neural network to model $\mathbf{q_\phi}(\mathbf{z}|\mathbf{x})$ and $\boldsymbol{p_\theta}(\boldsymbol{x}|\boldsymbol{z})$. The network includes both traditional fully-connected layers and the more recent convolutional and max-pooling layers and learns temporal dependencies as a window of the readings given at each time step. The convolutional and max-pooling layer enables the model to be invariant to small temporal translations (LeCun et al. 2015). A fully-connected layer has a connection between every neuron on the first layer to every neuron on the second layer, as seen in regular neural networks. A convolutional layer, on the other hand, applies a convolution operation to the input where in a single neuron in the first layer is connected to only a subset of the neurons in the second layer that are adjacent to it. A pooling layer pools a subset of the inputs of a layer together, and in particular a max-pooling layer pools the input by taking a max of the inputs. The number of inputs pooled by the max-pooling layer is a design parameter referred to here as the pooling size. An upscale layer is used in the decoder to reverse the max-pooling layers. Here this layer just replicates the input the same number of times as the pooling size. Also, after each fully connected layer and convolutional layer a Batch Normalization layer (Ioffe and Szegedy 2015) is added in order to improve the stability of the optimization process. A more elaborate description of the concepts can be found in a book by Goodfellow et al. (2016).

Adam optimization algorithm (Kingma and Ba 2014) is used to jointly train the encoder and the decoder by maximizing the variational lower bound. Adam is a stochastic gradient-descent optimization method for stochastic optimization that only needs first-order gradients with low memory requirement. An online learning model by design, VAE enables predictor update with streaming data, eliminating the need for training on the batch data every time new data becomes available.

This model is used here to learn the distribution of the normal system operation data. While the model training is relatively compute-intensive, its utilization thereafter for detection is very fast and computationally inexpensive. Let $\hat{\mathbf{x}}$ be the reconstruction of the input $\mathbf{x}$ produced by the model by passing the input through the encoder and then the decoder. The logarithm of the probability density of this reconstruction may be estimated as $\log(\mathbf{p_0}(\hat{\mathbf{x}}|\mathbf{z}) \times \mathbf{q_\Phi}(\mathbf{z}|\mathbf{x}))$. Thus, the logarithm of the probability density of the model to regenerate back the input can be estimated as $\log(\mathbf{p_0}(\hat{\mathbf{x}} = \mathbf{x}|\mathbf{z}) \times \mathbf{q_\Phi}(\mathbf{z}|\mathbf{x}))$. We define this as the log reconstruction probability function (LRP). If this value is too small, then the reading is flagged as an attack. This is because a low LRP value indicates that the correlations between the different inputs dimensions discovered by the model did not conform to that with the normal system operation data, suggesting an anomalous behavior. This also means the detection mechanism relies on a set threshold to binarize its judgement into a safe/attack state, but the continuous LRP values are still useful alongside to realize how confident the model is in its judgement.

## 3. Performance Evaluation Criteria

Performance of the model in detecting attacks is evaluated using confusion matrices, receiver operating characteristic (ROC) curves, and a metric that aggregates precision and recall.

An ideal detection model is able to flag all the true attacks and report no false alarms. The former is often called recall, quantified as the number of true attacks flagged by the model over the number of true attacks that exist. The latter is commonly referred to as precision, which is the ratio of the number of true attacks flagged by the model over the number of all the attacks it has flagged.

Precision and recall for an ideal detection model are both equal to one. In reality, they are often dependent and in conflict. $F_1$ score (Sokolova et al. 2006) is often used to aggregate the two competing metrics and come up with a single metric to evaluate the detection model:

$$F_1 = \frac{2 \times precision \times recall}{precision + recall} \tag{5}$$

Confusion matrices and ROC curves are also used in this study to evaluate the diagnostic performance of the model and its sensitivity to discrimination thresholds. ROC curves help illustrating the complete trade-off between true positive and false positive rates for a range of different LRP thresholds.

## 4. Data Requirements

At a minimum, the raw observations of the normal operation of the system provide the input for the anomaly detection. The raw data may be real or synthetic, depending on whether it is direct observations of a real system or a model of it, respectively. As an unsupervised learning technique, the proposed model does not require labeled attack data, except for test and evaluation purposes. In a water distribution system, this raw data generally includes the pump flows, tank levels, and large valve status. Detection can be bolstered by including auxiliary data such as pipe flow, pressure, and consumption data. Although not the focus of this work, the communication of the data (e.g., Cellular, propriety radio, mesh networks, etc.), which enables the ICS, introduces the vulnerability to cyberattacks. The purpose of this research is to algorithmically detect if the data received and the control rules set has been tampered with.

## 5. Experiment Setup

The cyberattack detection model developed in this study is versatile and suits broad sectors. In this application, it is demonstrated on two simulated cyberattack detection problems involving a water distribution network.

### *5.1. VAE Configuration*

The VAE model described here is built using the open-source framework Theano (Al-Rfou et al. 2016). Architecture of the VAE model developed and used in the two demonstration case studies is illustrated in Figure 2. The encoder half of the network consists of three pairs of convolutional and max-pooling layers followed by three dense layers. The decoder performs the reverse operation of the encoder and thus has three dense layers followed by three pairs of convolutional and upscaling layers. An upscale layer performs the reverse of a max pooling layer (Figure 2). The encoder half of the network consists of three pairs of convolutional and max-pooling layers followed by three dense layers. The decoder performs the reverse operation of the encoder and thus has three dense layers followed by three pairs of convolutional and upscaling layers. An upscale layer performs the reverse of a max pooling layer.

A rolling window marches through time in the 43 time series with one-hour steps and forms data points $x$, which are sequentially fed into the first layer of the neural recognition model. Each data point $x$, therefore, is 24 hours long and has 43 dimensions, and every two consecutive data points are one hour apart.

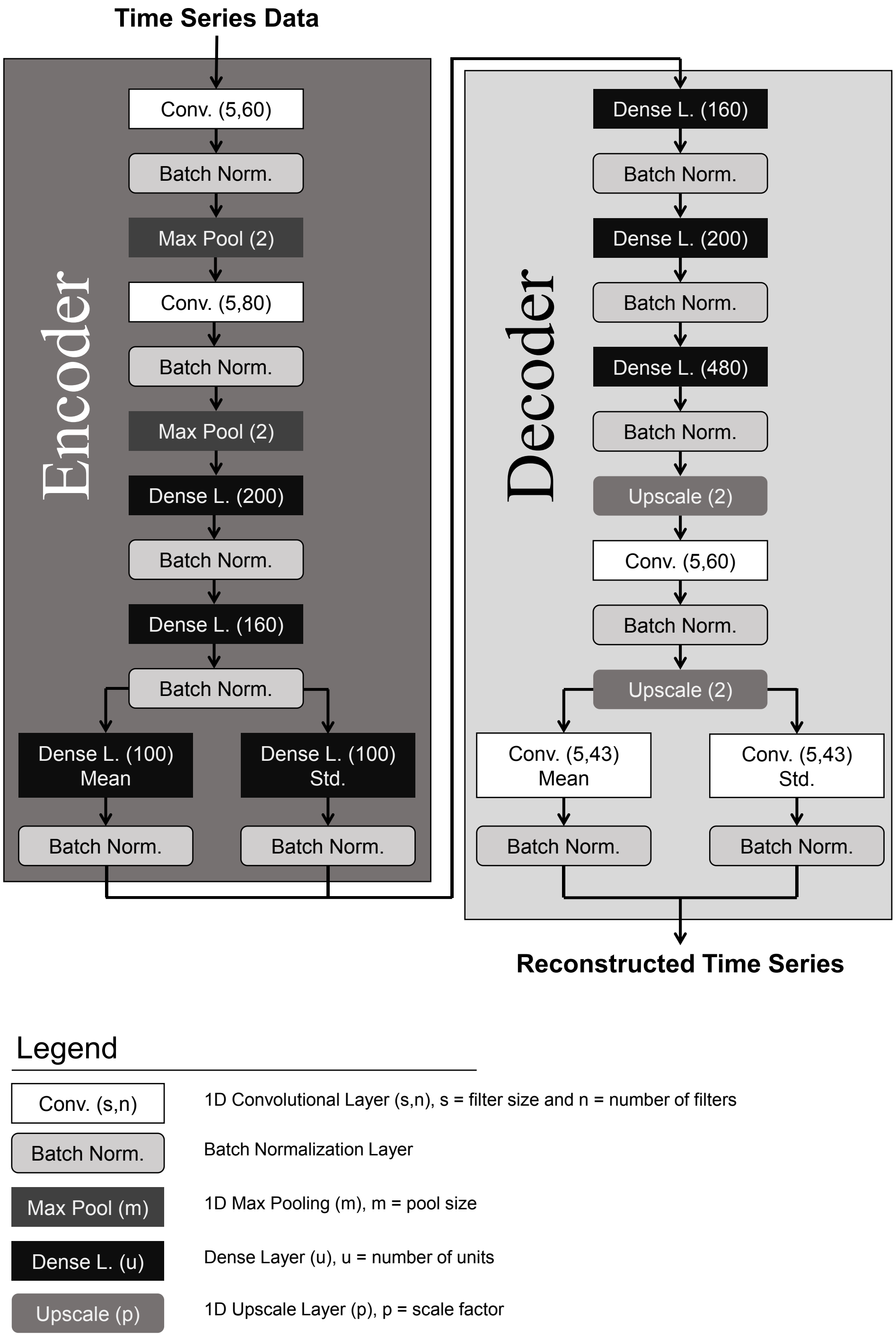

**Figure 2.** Architecture of the VAE model used in demonstration case studies

## 5.2. *Detection Threshold Setting*

Detection threshold setting is a critical and challenging task to accomplish and prevails in any detection problem and solver. It is critical in that an over-sensitive detection model leads into cry-wolf effect and an under-sensitive detection model results in a cyberattack left unnoticed. And it is challenging in that water distribution systems are dynamical systems with deep uncertainties and the risk introduced by cry-wolf effects and unnoticed cyberattacks are very difficult to quantify, if not impossible.

A primary output of the VAE model is LRP data, which defines a continuous attack likelihood space. It needs to be discretized into safe/attack flags for alarm automation purposes. However, it does not require only one threshold be used. In practice multiple thresholds could be utilized to represent varying levels of certainty. In this experiment, sensitivity of the model to the threshold setting is analyzed by a calculating the performance evaluation criteria described in Section 3 for various threshold settings. Optimal thresholds based on $F_1$ scores are also calculated here through enumeration. This performance-based threshold setting is dataset-dependent and requires availability of labeled data. If no actual or simulated attack data is available, one may alternatively set the threshold according to the statistical measures of the LRP values calculated for normal system operations time periods (e.g., quantile-based).

## 5.3. Hydraulic Network

The experiment involves C-Town water distribution network, a popular benchmark for water research projects, which is based on a real, medium-sized system (Ostfeld et al. 2011). C-Town has diverse SCADA rules, pump curves, and demand patterns, and thus suits the needs of this study. Network topology and feedback controls are illustrated in Figure 3.

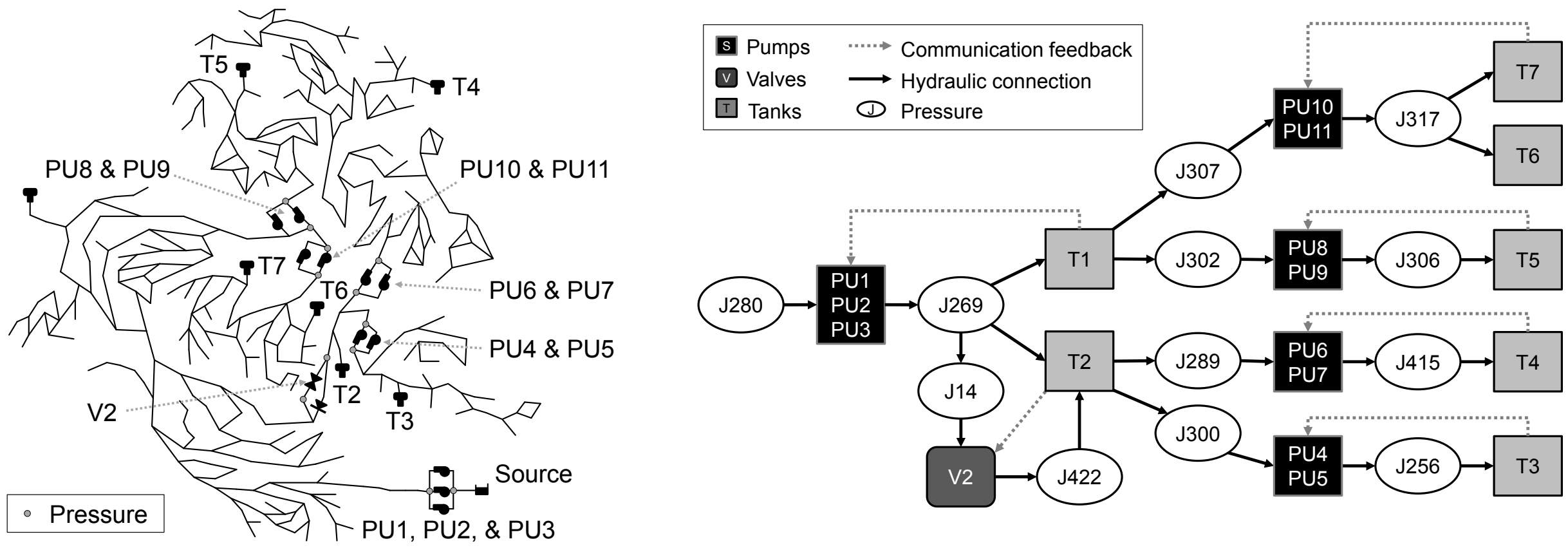

**Figure 3.** Network layout and hierarchy of the C-Town water distribution system. Pump status and flow, valve status and flow, pressure before and after pump stations, and tank levels are sensed by the SCADA. In total 43 system variables are collected in hourly intervals. The letter J stands for junction.

The model is demonstrated on two separate sets of training and test data both involving C-Town network, and will be referred to as Case 1 and Case 2. Data in Case 1 is an ensemble of training and attacks data that is generated in this article. Data in Case 2 is again an ensemble of training and attacks data it is generated by BATADAL (Taormina et al. 2018) using epanetCPA toolbox (Taormina et al. 2017) and is publicly available via www.batadal.net/data.html. Each Case includes a series of cyberattacks, which interfere with the performance of the C-Town network's components such as tanks, valves, and pumps. The only difference between the C-Town used in each case is that, in Case 1, a Gaussian noise is infused to all the nodal demands to partially account for the stochasticity in real world.

Both cases involve two datasets, each comprised of 43 hourly time series belonging to 43 sensed variables in the network (Figure 3). This include water levels for all the tanks (7 variables), inlet and outlet pressure for Valve V2 and all the five pumping stations (12 variables), and their status and flow (24 variables). The detailed description of each case is presented in the following two subsections.

### 5.4. Case 1

Case 1 uses data that is generated by this paper through extended-time hydraulic simulations. This also involves the use of a case-specific rule violation model, which is developed and used here for comparison purposes.

#### 5.4.1. Data Description

Two datasets are generated through extended-time hydraulic simulations with EPANET toolkit (Rossman 2000). The first dataset is one year long, includes no attacks, and is used for training purposes while the second one is two months long and contains attacks. To model an attack, the control statement, which is encoded and included in hydraulic file to change status or setting at some future point in the simulation, is manipulated using ENsetcontrol function in EPANET toolkit. This EPANET function allows altering the parameters of a simple control statement as the extended-time simulation is triggered. For example, this function is used to reconfigure a tank's water level threshold that triggers a feedback-controlled pump to turn on.

A Gaussian noise with mean of 0 and standard deviation of 0.02 times the point demand value is infused to all the nodal demands to model the real-world stochasticity to some extent.

The attack model used involves compromise of the feedback controllers. The attack dataset time series generated contains three separate, independent attacks listed in Table 1. Ample time gap is given between the attacks so that the disturbances generated by an attack are fully diminished before introduction of the next one. The attacks may harm the system if not promptly detected and mitigated. Attacks of type 1 and 2 may cause tank overflows. In this network, in particular, Attack 2 may cause Tank5 to overflow given the fact that the maximum level is 4.5 meters in C-Town. Attack 3 leads into pump short-cycling by leveling out an original 0.5-ft. buffer set between PU10 and PU11.

**Table 1.** Attack scenarios

| # | start time | duration (hour) | compromised control |
|---|---|---|---|
| 1 | 240 | 72 | PU8 Closed if Tank5 Level Above 4.3 (originally 4.0) ft. |
| 2 | 600 | 72 | PU8 Closed if Tank5 Level Above 5.0 (originally 4.0) ft. |
| 3 | 960 | 96 | PU11 Open if Tank7 Level Below 1.5 (originally 1.0) ft. |

#### 5.4.2. Rule Violation Model

The VAE model relies solely on SCADA data in its raw form, without a need for any data preprocessing or domain knowledge. For comparison, we also perform independently a detection process that simply checks for impossibilities in the data. First, tank levels are checked against the upper and lower limits of the tanks. These limits determine the lowest and highest level of water in a tank and are pre-configured by utilities for each tank. An anomaly is flagged when the SCADA reports values outside the level for each tank. Second, operational rules for pump stations are checked. Operational rules translate when pumps are on or off, based on tank levels. If the prescribed operational rule is not met, an anomaly is flagged. Third, a system wide mass balance is performed by checking that the volume stored is not greater than the volume pumped. This is done by converting tank levels to a volume using their diameters and integrating the change in level over time. Fourth, data are checked for hydraulic feasibility. We are not monitoring or using other pressures. Pressures at both sides of pump stations are converted to hydraulic head and hydraulic heads at the inlet side of nearby pump stations are compared to ensure the head values are similar. Within this comparison, the hydraulic head at nearby sensors that do not have pumping between them are compared. For example the inlet hydraulic head at the inlet to Pump PU9 is compared to Pump PU10. An anomaly is flagged when the downstream datum is higher than the upstream. Pump stations are hydraulically checked by comparing the difference in head between the intake and discharge, to the head added by the pump, which is calculated using the flow and pump curve. Anomalies are flagged when the head gains reported by the pressure readings do not align with the calculated gains suggested by the pump flow.

The result of each rule is binary, meaning identified anomalies are unambiguous since data reflect an impossibility. However, time steps in which no anomaly is detected do not preclude attacks since it is possible that an anomaly be operationally, physically, or hydraulically feasible. The rule inspection model is extremely precise with its true-positives, but not with its false-negatives, as will be also observed in the case studies later in the paper. For this reason this method is smoothed by flagging the previous 48 hours of data as suspicious whenever the rules find an anomaly. It represents a relaxation of the rules such that the scores are improved. One may, alternatively, label a time neighborhood after or around the rule violation hour (e.g., 24 before and 24 hours after) as attack or set a time neighborhood other than 48 hours. The objective here is not to present a perfect rule-based model but provide a realistic baseline score to compare with the variational autoencoder.

### 5.5. Case 2

Second case is taken from the literature and contains a more diverse set of attacks. This dataset used for evaluating the algorithm is from BATADAL (Taormina et al. 2018). Training dataset 1 (Taormina et al. 2018), which contains no attacks and represents normal operating conditions, is used for training the model. Test dataset (Taormina et al. 2018), which contains seven attacks, is used to test and evaluate the trained model. The attacks, which are thoroughly described in BATADAL (Taormina et al. 2018), are diverse and complex. In this study, the test dataset is renamed from Attacks 8 to 14 to Attacks 1 to 7, respectively. They involve alteration of the system operations through malicious activation of actuators and change of their settings, as well as deception attacks. They can be categorized in three ways: 1) changing thresholds so that the pump or valve operational rules are modified; 2) manipulating the tank level readings so that the pumps or valves operate improperly; and, 3) operating the pumps or valves independently of their operational rules. Additionally, these attacks can be concealed, meant to disguise the attack. The concealment methods are categorized as: 1) overwriting data that would indicate an attack with historic data; and 2) and offsetting the data so that the data appears within the normal operating range.

## 6. Results

The VAE model has been applied to the two cases. The results and discussions are described in this section.

### 6.1. Case 1

Figure 4 illustrates the output LRP values for Case 1. The lower the LRP for a data point x, the more confident the model that it is an attack. Manifestation of all the three attacks is clearly observable in the LRP plot confirming the ability of VAE in cyberattack detection in this problem.

Inspection for possible rule violations can flag Attacks 1 and 2, but fails to detect Attack 3. It is observed that both Attacks 1 and 2 are flagged early when the smoothed rule inspection model is used. This is due to the practice of merging the point violations detected by the smoothed rule inspection model after the attack starts. Therefore, this observation does not mean that the model is able to alarm early in a realtime implementation. Attack 3 involved turning on Pump 11 before Tank 7 drained to the proper level. This attack does not trigger a flag because the state is feasible while the tank is filling. Checking if the tank reached it minimum level before the pump is turned on is problematic due to the resolution of the data --- the tank could be filling for almost an hour between data reporting.

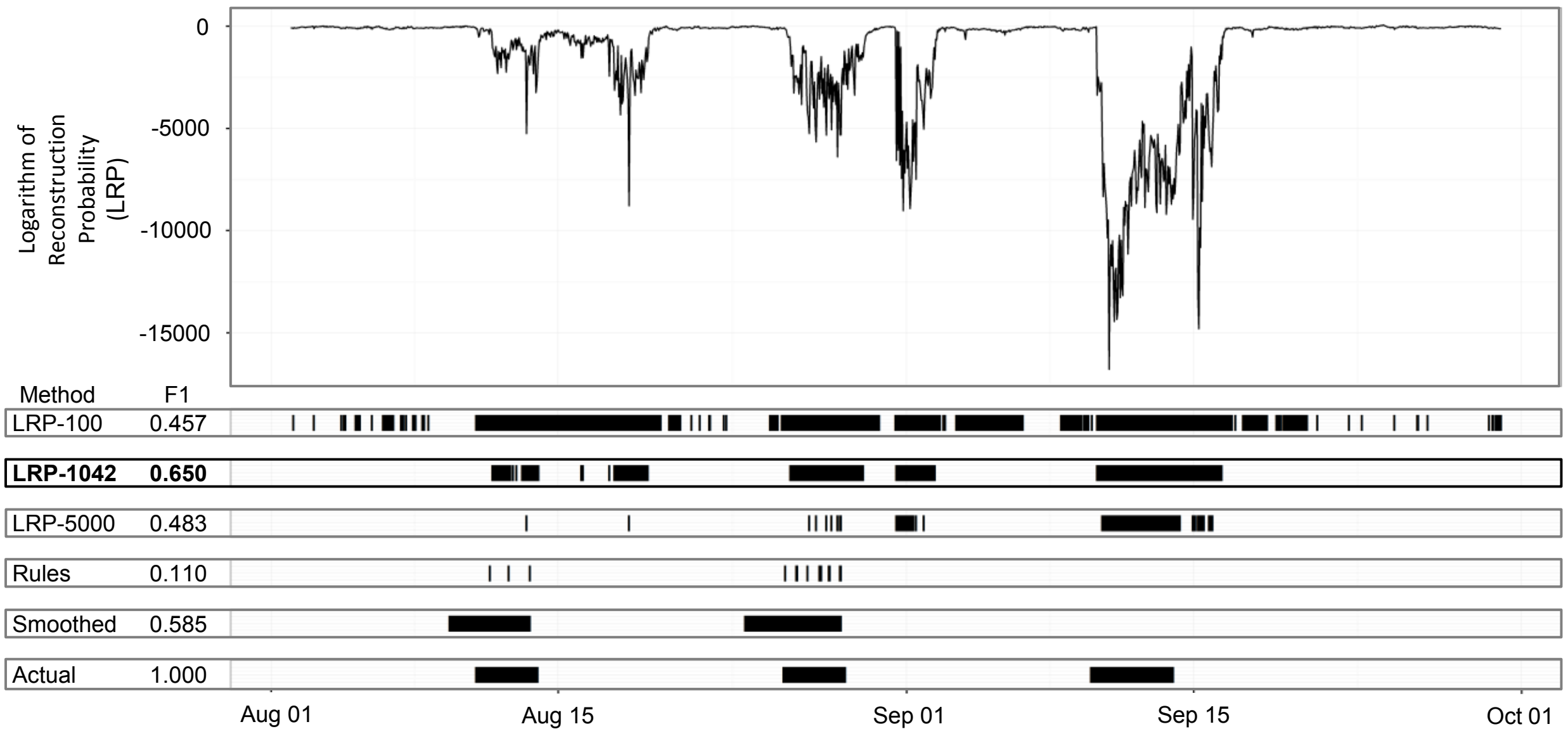

**Figure 4.** Logarithm of reconstruction probability (LRP) of data points in the attack dataset for Case 1. Detected attack labels associated with three LRP thresholds (-100, -1042, and -5000) are plotted against the actual labels and that from the original and smoothed rules inspection model.

For automation purposes, a threshold LRP can be set, below which an alarm will be triggered. A threshold of -1042 was found by enumeration to result in maximum $F_1$ score in this application, but selection of the threshold should be also informed by other objective and subjective measures and considerations. These can include, but not limited to, the sensors accuracy, the control automation level, the system vulnerability level, reports on any recent threats, and the severity of potential consequences. For example, the operator of a system that is deeply automated, relies on an old operating system, has been subject to recent phishing scams, or distributes water to critical services should be arguably more conservative and have a stricter threshold.

The threshold could also be set by running the model on a separate normal dataset and control for the number of false positives produced in it. In practice, cyberattacks are extreme and rare, therefore an operator would be more concerned with missing attacks than being interrupted by false alarms. This still does not mean that false alarm rate can be high, as this may develop mistrust into the detection model and lead to a cry-wolf effect. Moreover, inclusivity of a model in detecting the existence of attacks is the primary concern; its ability to accurately characterize attacks (such as their duration), while important, is a secondary concern.

A secondary drop in LRP is consistently observed in Figure 4 after any of the three attacks end (i.e., its comprised control rule is reset to its original state). This observation may be arguably due to the disturbance introduced and propagated by the act of terminating an attack. This state of abnormality is transitory, meaning the system returns to normal shortly as it has manifested itself as high LRP values. But until it fully diminishes, it is flagged by the VAE model and degrades its $F_1$ score. One may argue that this transitory state of the system should be still regarded as the attack state and any detection therein be counted as true positives. Because even though the attack is terminated, its effects are still prevailing. To evaluate this secondary disturbance, an experiment is conducted by simulating two hydraulic models with the same random seed for generating random numbers in the attack model. An attack is simulated in one of the models and compared with the other model, which has no attack. The comparison reveals that the water network requires some time steps to attain its normal performance after the attack is terminated.

The described argument that the transitory state of the system after should be still regarded as the attack state is a reasonable, alternate perspective, but here only the exact time period where a control rule is compromised is considered as the attack. The manifestation of this secondary disturbance can provide the operator with an insight into the period during which the system is still under the influence of a now neutralized attack. This additional information can be helpful for response and recovery purposes as well as understanding and enhancing resilience of the system against different attack scenarios. This idea of a post-attack disturbance being responsible for the secondary drop in LRP can be cross-examined by a comparable but different statistical technique that can similarly discern changes in a large set of heterogeneous time series (Aminikhanghahi and Cook 2017). Such a statistical analysis may also enable investigation of the statistical similarity of the no-attack and attack trajectories.

Beyond the $F_1$ score, a confusion matrix and ROC curve are used to evaluate the thresholds. The confusion matrix, Figure 5, allows the direct comparison between each predicted and actual condition. The threshold LRP-1042 balances the conditions in the confusion matrix and falls favorably on the ROC curve on false positive rate of 0.13 and true positive rate of 0.79 (Figure 6).

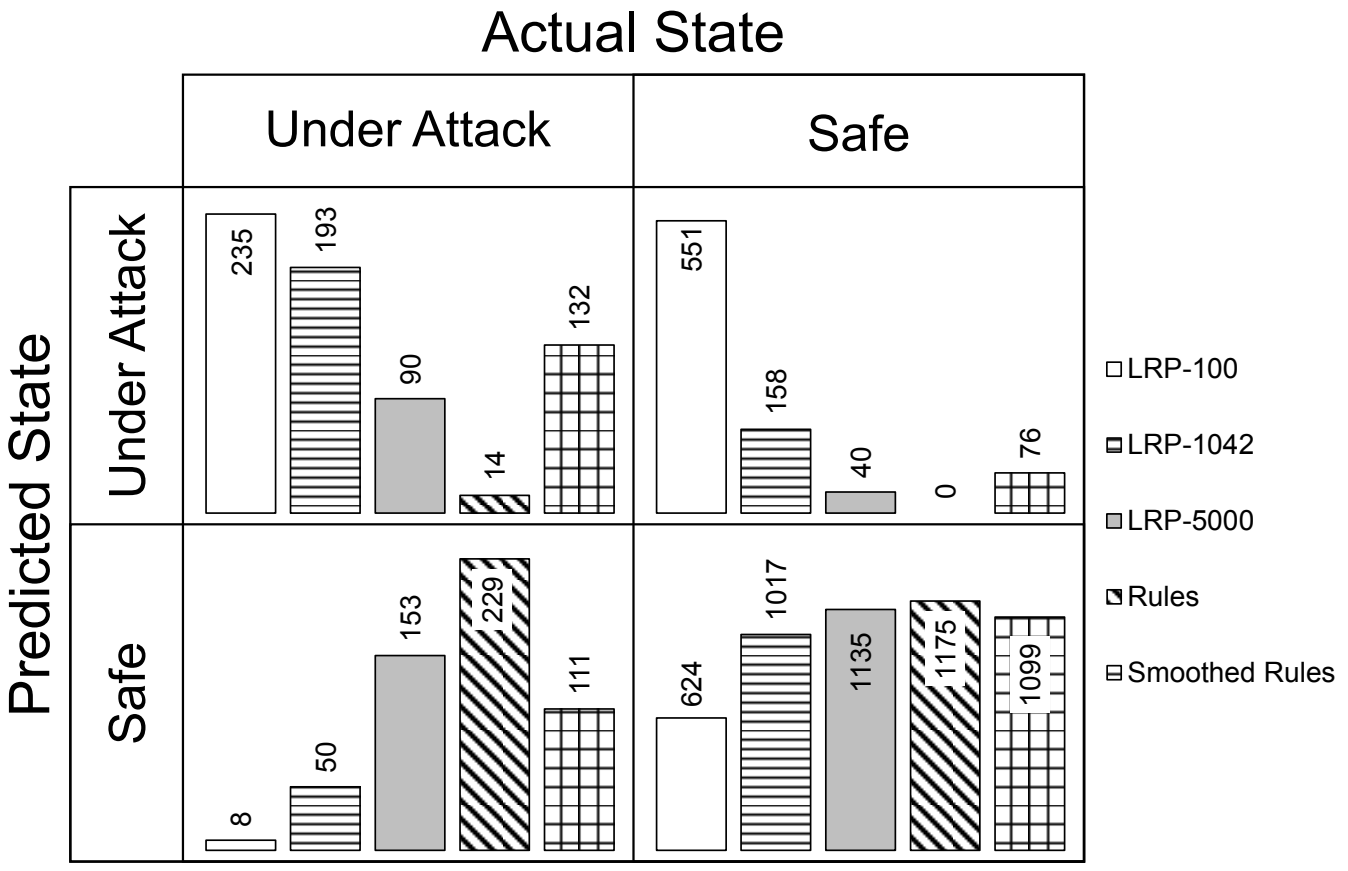

**Figure 5.** Confusion matrix for Case 1

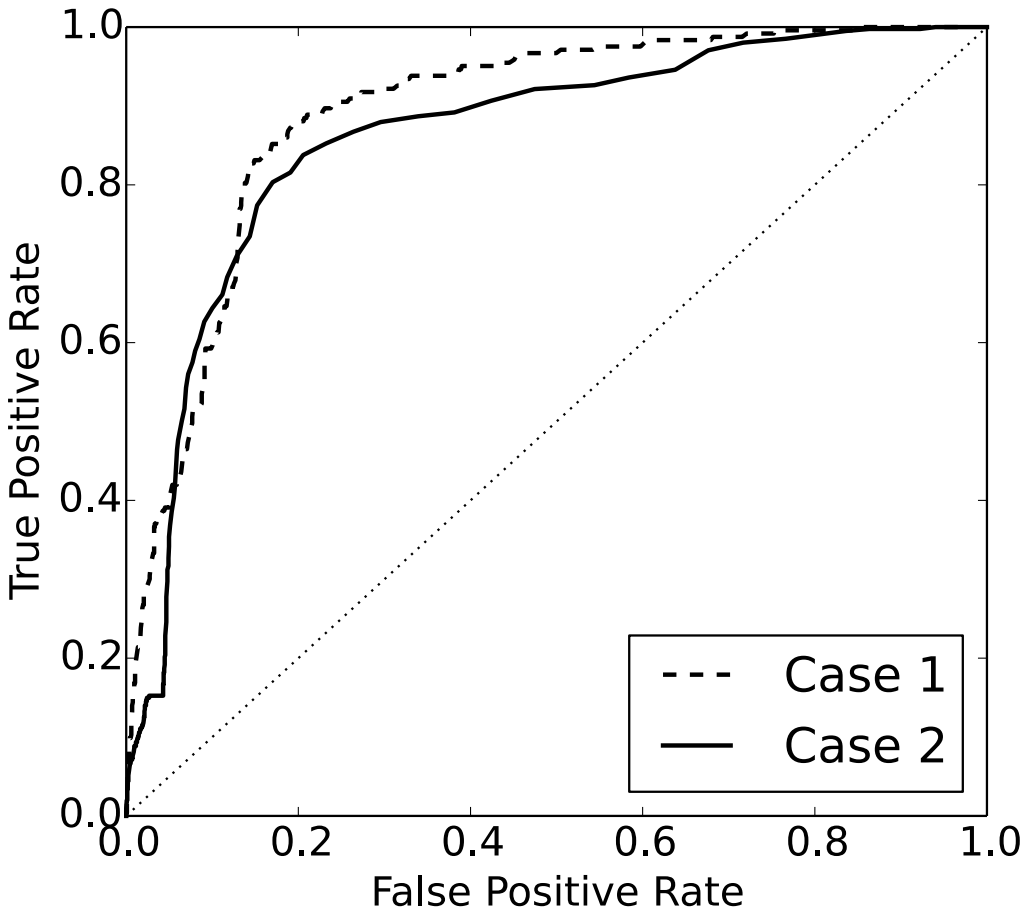

**Figure 6.** ROC curves for Cases 1 and 2. The area under the ROC curves for Case 1 and 2 are approximately 0.89 and 0.86, respectively (the area under an ideal ROC curve is 1.00)

## *6.2. Case 2*

LRP map for Case 2 is illustrated in Figure 7. Attacks 1 to 5 are distinctly manifested in the LRP curve and are easily distinguishable. These attacks are flagged by the model even at the relatively insensitive (small) discrimination threshold of -1000.

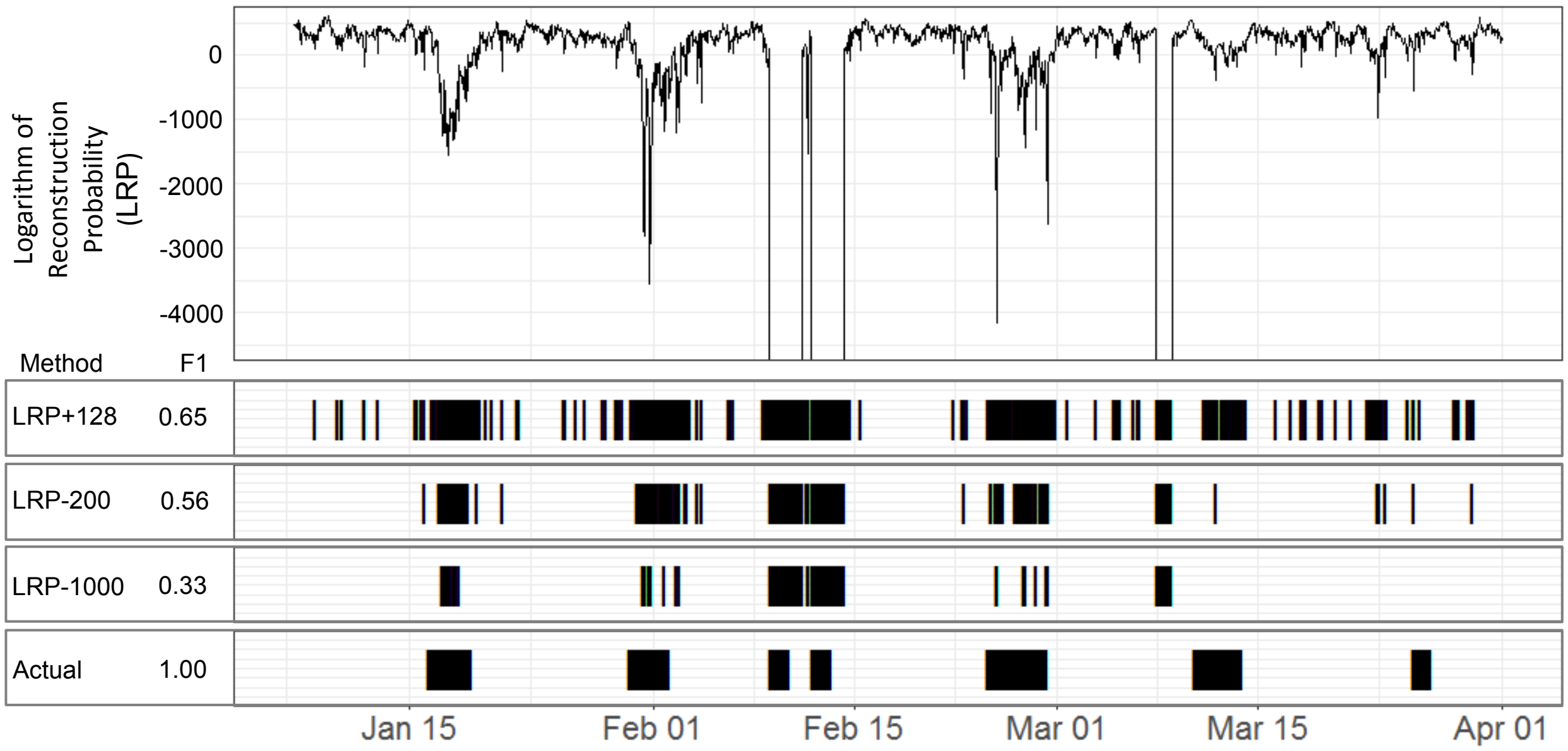

**Figure 7.** Logarithm of reconstruction probability (LRP) of data points in the attack dataset for Case 2. Detected attack labels associated with three LRP thresholds (+128, -200, and -1000) are plotted against the actual labels

Attacks 6 and 7 emerge in the detection as more strict thresholds are set, starting approximately at LRP -200. But it is achieved at the cost of false positives appearing alongside, harming precision of the detection. This can be observed in confusion matrix shown in Figure 8 as well.

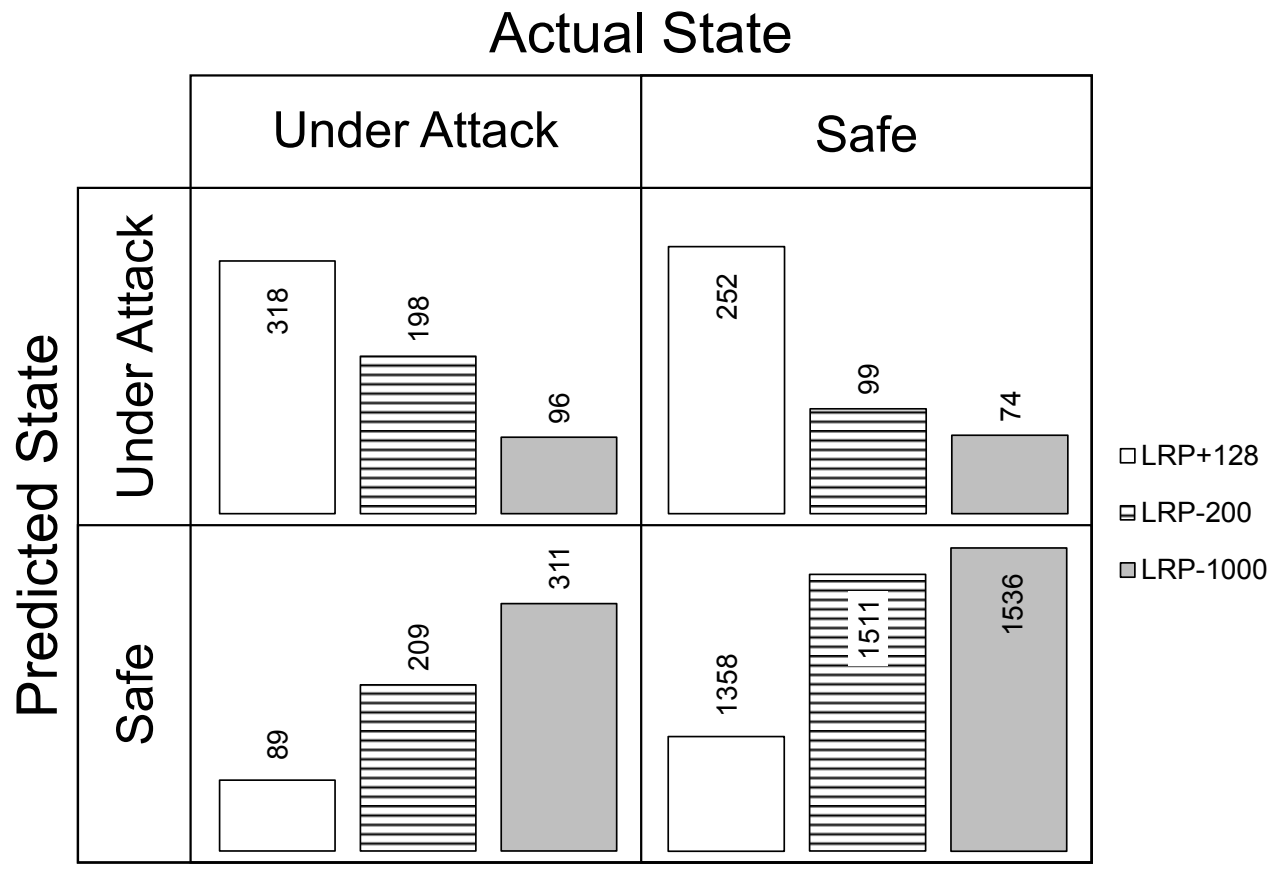

**Figure 8.** Confusion matrix for Case 2

According to BATADAL (Taormina et al. 2018), Attack 6 is created by two simultaneous actions: an attack and a concealment. The concealment ensures the SCADA system would not reflect the shortage of water in Tank 4, making detection of the attack very challenging. Such attacks might be detected better, if the time series for sub-regions of the network are monitored in isolation, but this involved the added burden of making manual settings and embedding expert knowledge.

It is observed that there exit a false positive before Attack 6, starting on March 8th 00:00 and lasting for one day, which persists to exit at the relatively insensitive LRP threshold of -1000 and even much lower thresholds. It is a curious case because this is the only time in all the data for both Cases 1 and 2 where the model has confidently flagged a window but has not been an actual attack. A manual investigation of the time series revealed that it has been due to a very abnormal operational behavior that happened at that time. Pump PU6 is activated on March 7th 23:00 for the first and only time in the three months of the test data, and then goes off again after three hours. An investigation into the training data reveals that PU6 has been active for a total duration of only 16 hours during the total 8760 hours of the available training data, which is translated into a 0.0018 occurrence probability. This event, while safe, is an obvious anomalous condition. Training the model on longer and more heterogeneous, inclusive observations data may alleviate the possibility of such false alarms. Right labeling of such extremely rare, but safe, conditions, nevertheless, is believed to remain a challenge for any anomaly-based detection system. It is one of the possible scenarios where rule-based approached can come to aid and complement the detection platform to reduce false positives. Such safe conditions, despite being rare, are often planned and designed conditions, and, thus, may not violate operational rules of the system when they occur. In this case of abnormal activation of PU6 on March 7th and the associated false positive alarm triggered, for example, an investigation of the training data and C-Town hydraulic file reveals that it is PU7 that very predominantly supplies water to the area covered by this PU6-PU7 pump station.

The threshold values associated with maximum $F_1$ was found to be different for Case 1 and Case 2. This may be attributed to the differences in the two systems that generated the simulated data for these two cases. As explained in Section 5, while the two cases share the same hydraulic network, the demands set for each case have been set differently. The demands in real world are determined by the system and this issue is thus not present. Adaptive threshold strategy (Ghafouri et al 2016) may also be used to account for seasonal demands patterns shifts in real systems.

According to the area under the ROC curves in Figure 6 and the confusion matrix in Figures 5 and 8, the VAE performs comparatively similar for both cases. The area under the ROC curve is improved 0.03 for Case 1. Additionally, a visual comparison of the LRP plots indicates that the attacks in Case 1 have manifested more distinctively compared to the no-attack time periods. This marginally better performance may be due to the higher diversity and complexity of the attacks in Case 2 as well as the higher inclusivity of the Case 1 data, which has been infused by using comparatively more variable demands during simulations.

## 7. Concluding Remarks

A cyberattack detection model was developed and demonstrated. The model relies solely on sensor reads data in their raw form and requires no preprocessing, system knowledge, or domain expertise to function. It is generic and can be readily applied to a broad array of ICS's in various industry sectors. Nevertheless, it is not perfect and has its own requirements (e.g., availability of vast amount of system observations data) and drawbacks (e.g., sensitivity to rare but planned operations such as activation of emergency booster pumps), as explained throughout the manuscript. Several methodological improvements are described below for future studies.

The methodology was applied here to a computer-simulated detection problem involving a water distribution system. While this offers several advantages (e.g., the availability of all labels and the possibility of conducting sensitivity and what-if analyses), future demonstrations on physical testbeds (Goh et al. 2016) or real-world systems are still necessary to account for the facets (e.g., the type variety of cyberattacks) not adequately captured in simulated problems.

Model-based and data-driven approaches each have distinct requirements, their own merits and drawbacks, and offer trade-off in complexity, performance, and costs. Model-based algorithms (such as Amin et al. (2013a, b) and Housh and Ohar (2017)) can be loosely coupled with the data-driven VAE to improve the detection performance, specially correcting false positives triggered by extremely rare but safe scenarios.

As an anomaly detection technique, the proposed VAE model offers the ability to detect previously-unknown attacks (also known as zero-day attacks). However, because any divergence from the normal patterns is flagged whether or not it is due the real threat, a high false-alarm rate is deemed possible. Providing the model with richer data and pairing it with a misuse detection model would help alleviating this problem.

The binary detection alarms generated by the VAE algorithm are sensitive to the chosen detection threshold. Clearly, a relatively sensitive threshold (large LRP) can create a larger number of false positives as discussed earlier. This study proposed the application of the VAE for attack likelihood, performed sensitivity analyses on the effects of the threshold chosen, and calculated optimal thresholds through a basic enumeration process based on the $F_1$ score. A simple way to combat the difficulty of threshold setting is by using multiple thresholds, representing varying degrees of confidence. Advanced threshold setting models such as Laszka et al. (2016) and Ghafouri et al. (2016) can be coupled with the proposed VAE model in future studies for optimal threshold setting. Ghafouri et al. (2016) used a game-theoretic setup to solve the problem of finding time-varying optimal detection thresholds in dynamical environments, with a numerical demonstration on water distribution systems. The new mechanism can also use a secondary set of data (e.g., the BATADAL Training Dataset #2, which is found in Taormina et al. 2018) for evaluation and adjustment of the optimal thresholds found.

The current VAE model presumes that data points are independent and identically distributed random variables. A relaxation of this presumption is expected to boost the model's approximation flexibility, hence augmenting its detection sensitivity and specificity. Both the binary variable and continuous variable inputs were modeled as Gaussian by the current model. Modeling the binary variables as a Bernoulli distribution should further improve the result. Efforts are underway to introduce these flexibilities.

## 8. Acknowledgements

Author contributions: S.E.C. formulated the methodology, performed analysis, and assisted with manuscript writing; A.R. orchestrated the work, wrote the manuscript, and assisted with analysis; Z.A.B formulated the comparison method and assisted with analysis; M.E.S. designed the attack generator and assisted with analysis.

## 9. References

Ahmed, C. M., Murguia, C., & Ruths, J. (2017). Model-based attack detection scheme for smart water distribution networks. In *Proceedings of the 2017 ACM on Asia Conference on Computer and Communications Security* (pp. 101-113).

Al-Rfou, R., Alain, G., Almahairi, A., Angermueller, C., Bahdanau, D., Ballas, N., ... & Bengio, Y. (2016). Theano: A Python framework for fast computation of mathematical expressions. *arXiv preprint*: https://arxiv.org/pdf/1605.02688.pdf.

Amin, S., Litrico, X., Sastry, S. S., & Bayen, A. M. (2013b). Cyber security of water SCADA systems—Part II: Attack detection using enhanced hydrodynamic models. *IEEE Transactions on Control Systems Technology*, 21(5), 1679-1693.

Amin, S., Litrico, X., Sastry, S., & Bayen, A. M. (2013a). Cyber security of water SCADA systems—Part I: Analysis and experimentation of stealthy deception attacks. *IEEE Transactions on Control Systems Technology*, 21(5), 1963-1970.

Aminikhanghahi, S., & Cook, D. J. (2017). A survey of methods for time series change point detection. *Knowledge and information systems*, 51(2), 339-367.

Bai, Y., Chen, Z., Xie, J., & Li, C. (2016). Daily reservoir inflow forecasting using multiscale deep feature learning with hybrid models. *Journal of Hydrology*, 532, 193-206.

Bengio, Y., Courville, A., & Vincent, P. (2013). Representation learning: A review and new perspectives. *IEEE transactions on pattern analysis and machine intelligence*, 35(8), 1798-1828.

Blei, D. M., Kucukelbir, A., & McAuliffe, J. D. (2017). Variational inference: A review for statisticians. *Journal of the American Statistical Association*, 112(518), 859-877.

Cardenas, A., Amin, S., Sinopoli, B., Giani, A., Perrig, A., & Sastry, S. (2009). Challenges for securing cyber physical systems. In *Workshop on future directions in cyber-physical systems security* (Vol. 5).

Chandola, V., Banerjee, A., & Kumar, V. (2009). Anomaly detection: A survey. *ACM Computing Surveys*, 41(3), 15.

Ghafouri, A., Abbas, W., Laszka, A., Vorobeychik, Y., & Koutsoukos, X. (2016). Optimal thresholds for anomaly-based intrusion detection in dynamical environments. In *International Conference on Decision and Game Theory for Security* (pp. 415-434). Springer, Cham.

Goh, J., Adepu, S., Junejo, K. N., & Mathur, A. (2016) A Dataset to Support Research in the Design of Secure Water Treatment Systems. *The 11th International Conference on Critical Information Infrastructures Security*.

Goodfellow, I. J., Pouget-Abadie, J., Mirza, M., Xu, B., Warde-Farley, D., Ozair, S., Courville, A. & Bengio, Y. (2014, December). Generative adversarial networks. In *Proceedings of the 27th International Conference on Neural Information Processing Systems* (pp. 2672-2680). MIT Press.

Goodfellow, I., Bengio, Y., & Courville, A. (2016). Deep learning. MIT Press, Cambridge, MA.

Gregor, K., Danihelka, I., Mnih, A., Blundell, C., & Wierstra, D. (2014). Deep AutoRegressive Networks. In *International Conference on Machine Learning* (pp. 1242-1250).

Hinton, G. E., & Ghahramani, Z. (1997). Generative models for discovering sparse distributed representations. *Philosophical Transactions of the Royal Society of London B: Biological Sciences*, 352(1358), 1177-1190.

Housh, M., & Ohar, Z. (2017). Model based approach for cyber-physical attacks detection in water distribution systems. In *World Environmental and Water Resources Congress 2017* (pp. 727-736).

ICS-CERT (2016). NCCIC/ICS-CERT year in review: FY 2015. Report No. 15-50569, *U.S. Department of Homeland Security*, Washington, D.C.

Kingma, D., & Ba, J. (2015). Adam: A method for stochastic optimization. *3rd International Conference on Learning Representations (ICLR 2015),* San Diego, United States.

Kingma, D., & Welling, M. (2014). Auto-encoding variational Bayes. *2nd International Conference on Learning Representations (ICLR 2014)*, Banff, Canada.

Krizhevsky, A., Sutskever, I., & Hinton, G. E. (2012). ImageNet classification with deep convolutional neural networks. *25th International Conference on Neural Information Processing Systems (NIPS 2012*), Granada, Spain.

Laszka, A., Abbas, W., Sastry, S. S., Vorobeychik, Y., & Koutsoukos, X. (2016). Optimal thresholds for intrusion detection systems. *In Proceedings of the Symposium and Bootcamp on the Science of Security* (pp. 72-81). ACM.

Laszka, A., Abbas, W., Vorobeychik, Y., & Koutsoukos, X. (2017). Synergic security for smart water networks: redundancy, diversity, and hardening. In *Proceedings of the 3rd International Workshop on Cyber-Physical Systems for Smart Water Networks* (pp. 21-24). ACM.

LeCun, Y., Bengio, Y., & Hinton, G. (2015). Deep learning. *Nature*, 521(7553), 436-444.

Marçais, J., & de Dreuzy, J. R. (2017). Prospective interest of deep learning for hydrological inference. *Groundwater*, 55(5), 688-692.

Ostfeld, A., Salomons, E., Ormsbee, L., Uber, J. G., Bros, C. M., et al. (2011). Battle of the water calibration networks. *Journal of Water Resources Planning and Management*, 138(5), 523-532.

Perelman, L., & Amin, S. (2014). A network interdiction model for analyzing the vulnerability of water distribution systems. In *Proceedings of the 3rd international conference on High confidence networked systems* (pp. 135-144). ACM.

Rao, V. M., & Francis, R. A. (2015). Critical review of cybersecurity protection procedures and practice in water distribution systems. In *IIE Annual Conference. Proceedings* (p. 2019). Institute of Industrial and Systems Engineers (IISE).

Rasekh, A., Hassanzadeh, A., Mulchandani, S., Modi, S., & Banks, M. K. (2016). Smart water networks and cyber security. *Journal of Water Resources Planning and Management*, 01816004.

Rossman, L. A. (2000). EPANET 2: users manual. Environmental Protection Agency, Washington, DC.

Sergey Ioffe and Christian Szegedy (2015). Batch Normalization: Accelerating Deep Network Training by Reducing Internal Covariate Shift, Proceedings of Machine Learning Research, 37, 448—456.

Sokolova, M., Japkowicz, N., & Szpakowicz, S. (2006). Beyond accuracy, F-score and ROC: a family of discriminant measures for performance evaluation. In *Australian Conference on Artificial Intelligence* (Vol. 4304, pp. 1015-1021).

Stouffer, K., Falco, J., & Scarfone, K. (2011). Guide to industrial control systems security. *NIST special publication*, 800(82), 16-16.

Tao, Y., Gao, X., Hsu, K., Sorooshian, S., & Ihler, A. (2016). A deep neural network modeling framework to reduce bias in satellite precipitation products. *Journal of Hydrometeorology*, 17(3), 931-945.

Taormina, R., Galelli, S., Tippenhauer, N. O., Salomons, E., & Ostfeld, A. (2017). Characterizing cyber-physical attacks on water distribution systems. *Journal of Water Resources Planning and Management*, 04017009.

Taormina, R., Galelli, S., Tippenhauer, N. O., Salomons, E., & Ostfeld, A., et al. (2018). The battle of the attack detection algorithms: disclosing cyber attacks on water distribution Networks. Journal of *Water Resources Planning and Management*, DOI: 10.1061/(ASCE)WR.1943-5452.0000969.

Thomson, L. L. (2016). Insecurity of the internet of things. Scitech Lawyer, 12(3), 32.

Verizon Communication (2016). Data breach digest: scenarios from the field. New York, NY.

Water Sector Coordinating Council (2008). Roadmap to Secure Control Systems in the Water Sector. *National Association of Water Companies*, Washington DC.

White House (2013). Presidential Policy Directive–Critical Infrastructure Security and Resilience. PPD-21. Washington, DC.

White House (2017). Presidential Executive Order on Strengthening the Cybersecurity of Federal Networks and Critical Infrastructure. Washington, DC.